\documentclass[aps,prl,twocolumn,superscriptaddress,nofootinbib,floatfix,noshowpacs]{revtex4-1}
\pdfoutput=1
\newif\ifcomment
\usepackage{dcolumn}
\usepackage{bm}
\usepackage{amsmath,amssymb,amscd,amsfonts}
\usepackage{color,hyperref,url}
\usepackage{listings}
\usepackage{slashed}
\usepackage{color}
\usepackage[pdftex]{graphicx}
\usepackage{epstopdf}
\usepackage{epsfig}
\usepackage{grffile}
\usepackage{relsize}
\usepackage{multirow}
\usepackage{xspace}
\usepackage{floatrow}
\usepackage{calrsfs}
\usepackage{txfonts}
\graphicspath{{./img/}}
\let\OLDthebibliography\thebibliography
\renewcommand\thebibliography[1]{
  \OLDthebibliography{#1}
  \setlength{\parskip}{0pt}
  \setlength{\itemsep}{5pt plus 0.3ex}
}

\newcommand{\com}[1]       {\relax}

\newcommand{\PbPb}         {\mbox{Pb--Pb}}

\newcommand{\snn}          {\ensuremath{\sqrt{s_{\mathrm{NN}}}}}

\newcommand{\Tab}[1]       {Tab.~\ref{#1}}

\newcommand{\Figure}[1]    {Figure~\ref{#1}}

\newcommand{\Ref}[1]       {Ref.~\cite{#1}}

\newcommand{\pT}           {\ensuremath{p_{\mathrm{T}}}}

\newcommand{\jetraw}       {\ensuremath{{p}_{\mathrm{T,\;ch\,jet}}^{\mathrm{raw}}}}
\newcommand{\jetrec}       {\ensuremath{{p}_{\mathrm{T,\;ch\,jet}}^{\mathrm{rec}}}}
\newcommand{\jettrue}      {\ensuremath{{p}_{\mathrm{T,\;ch\,jet}}^{\mathrm{true}}}}

\newcommand{\GeVc}         {GeV/$c$}

\begin{document}
\title{Machine Learning based jet momentum reconstruction in heavy-ion collisions}
\author{R\"udiger Haake}
\affiliation{Yale University, Wright Laboratory, New Haven, CT, USA}
\author{Constantin Loizides}
\affiliation{ORNL, Physics Division, Oak Ridge, TN, USA}
\date{\today}
\begin{abstract}\noindent 
The precise reconstruction of jet transverse momenta in heavy-ion collisions is a challenging task.
A major obstacle is the large number of (mainly) low-\pT\ particles overlaying the jets.
Strong region-to-region fluctuations of this background complicate the jet measurement and lead to significant uncertainties.
In this paper, a novel approach to correct jet momenta (or energies) for the underlying background in heavy-ion collisions is introduced.
The proposed method makes use of common Machine Learning techniques to estimate the jet transverse momentum based on several parameters, including properties of the jet constituents.
Using a toy model and HIJING simulations, the performance of the new method is shown to be superior to the established standard area-based background estimator.
The application of the new method to data promises the measurement of jets down to extremely low transverse momenta, unprecedented thus far in data on heavy-ion collisions.
\end{abstract}
\maketitle

\section{Introduction}
In ultrarelativistic heavy-ion collisions a new state of nuclear matter is created: the Quark-Gluon Plasma (QGP)~\cite{Bhattacharya:2014ara}.
In the QGP, deconfined quarks and gluons interact strongly and form a hot and dense medium that can be approximately described by hydro- and thermodynamics.
The regime of strong coupling at large distances, especially in systems of high temperature or large energy densities, is still not well understood in Quantum Chromodynamics (QCD).
An ideal self-generated probe to explore the properties of the medium and its interactions are particle jets~\cite{Gyulassy:1990ye}.
Reconstructed in the detector as collimated sprays of color-neutral particles, jets are created in a large-momentum-transfer scattering of partons in the early stage of a high-energy collision.
Their production is well understood within the framework of QCD and their rates can be perturbatively calculated in vacuum.
In a heavy-ion collision, jets traverse the strongly-interacting medium and interact mainly non-perturbatively and, thus, can serve as valuable probes of the QGP.

The reconstruction of particle jets in heavy-ion collisions is a complex task. The main obstacle is the overwhelmingly large background of particles that do not originate from hard interactions.
In ALICE~\cite{Aamodt:2008zz}, the mean momentum density in 0--10\% most central collisions at $\snn =2.76$ TeV leads to a contribution to the jet momentum that is already of the order of the typical jet momentum itself. The average charged particle transverse momentum density for particles with momenta above $0.15$ \GeVc\ is $\langle \rho \rangle \approx 138.2$ \GeVc\ per unit area, while its standard deviation is $\sigma(\rho) = 18.5$ \GeVc\ ~\cite{Abelev:2012ej}. Since jets are rare objects, these numbers provide already a good estimate of the mean background in the selected events.
In addition, this background shows large uncorrelated and also correlated region-to-region fluctuations. Uncorrelated fluctuations are due to random Poissonian fluctuations of the number of particles and their momenta. Sources of correlated fluctuations are e.g.\ physical correlations of the particles from the particle flow or also the non-uniform detector acceptances.
These fluctuations have a large impact on the reconstructed jet momentum and on the jet axes by directly affecting the jet finding algorithm and eventually result in large uncertainties on the final measurements.
An approach to at least lower the impact of the background at the expense of a potential fragmentation bias is a higher $\pT$-cut for constituents used in the jet finding algorithm. This massively reduces the background, which mostly consists of low-$\pT$ particles, but it also discards the low-$\pT$ parts of the jet.
The treatment of the background and its fluctuations depends on the observable under study. In this paper, the focus is on the correction of observables based on jet momentum, i.e.\ the correction of the jet energy scale, without applying a particular constituent cut. The impact of the background on the jet (sub)structure, e.g.\ by distorting the jet axis, is not discussed here.

In the standard method for jet spectra measurements in ALICE, the background momentum density per unit area is calculated on an \textit{event-by-event} basis. Each jet is then corrected by taking into account the event-averaged density multiplied by the jet area.
The area-based method corrects the jet momentum for the average background but leads to large residual fluctuations.
These residual fluctuations are then typically corrected for on a statistical basis in an unfolding procedure, see for instance~\Ref{Abelev:2013kqa}.

The new approach, introduced in this paper, calculates the corrected jet momentum on a \textit{jet-by-jet} basis to reduce the residual fluctuations and to allow a more precise estimate for the jet momentum. As we demonstrate below, this enables the measurement of jets in heavy-ion collisions with transverse momenta much lower than what is currently possible.
We apply Machine Learning~(ML) techniques, which are widely used in the HEP community~\cite{Albertsson:2018maf},
to obtain the mapping between jet parameters, e.g.\ constituent momenta, and the \textit{true} transverse momentum of the jet.
The background estimator is trained on a toy model that embeds jets with known (true) transverse momenta in a simulated thermal heavy-ion background.
Using this toy model, the performance of the new method is compared to that of the established correction method.
To demonstrate the applicability of the new approach on jet populations in heavy-ion collisions, the correction method performance is then also presented when applied onto HIJING~\cite{hijing} simulations.

\section{Jet definition and toy datasets}
The basis of this analysis are reconstructed charged jets.\footnote{An extension of this method to fully-reconstructed jets including neutral particles is straightforward.}
They are reconstructed by the anti-$k_\mathrm{T}$ algorithm~\cite{Cacciari:2008gp} implemented in Fastjet~\cite{Cacciari:2011ma}.
Since the background and its fluctuations increase quadratically in $R$, current jet analyses in heavy-ion physics typically avoid larger resolution parameters.
The resolution parameter is chosen deliberately at the relatively high value $R = 0.4$ to test the method.
Charged particles above $\pT = 0.15$ GeV/$c$ are taken into account in the jet reconstruction.\\

In contrast to an algorithmic correction approach where the correction, i.e.\ the mapping of raw and corrected jet momentum, is directly described, ML techniques \textit{learn} the mapping from a training dataset.
The quality of the training dataset is crucial for the final performance of the correction method on real data.
It should only encode well-known physics features and should be as simple as possible.
A good dataset allows the method to learn general features connected to the problem and enables the method to generalize to even slightly different data.
To judge its performance, the correction method is evaluated on the toy dataset and directly compared to the established correction method (described below).
The events in the toy dataset used for the evaluation are not used in the training phase.\\

To create events with particle jets in a heavy-ion background, PYTHIA-generated events are embedded in a thermal background.
The events from PYTHIA~\cite{Sjostrand:2006za} are generated at $\sqrt{s} = 2.76$ TeV with particles in a pseudorapidity range of $|\eta| < 0.9$ and in full azimuth.

The PYTHIA part of the events is created with PYTHIA6, using the Perugia 2011 tune~\cite{Skands:2010ak}. The thermal background is created by randomly distributing charged particles according to realistic particle multiplicity and momentum distributions. The multiplicity distribution is modeled with a Gaussian function with a mean of 1800 and a width of 200. This roughly reproduces the multiplicity in central \PbPb\ events at $\snn = 2.76$ TeV~\cite{Aamodt:2010pb,Aamodt:2010cz}.
The impact of different thermal background multiplicities on the model performance will be evaluated below.
The momentum distribution is a modified power law function and defined such that it coincides with the momentum distribution at low $\pT$~\cite{Aamodt:2010jd} but falls much faster for higher particle momenta roughly above 4 \GeVc.
Actual details on the higher-$\pT$ region of the thermal spectrum do not influence the model much, as this region only corresponds to a small part of the dataset.

For the actual application of the correction method to real detector-level heavy-ion events, the method should be trained on a dataset that describes the real, reconstructed data as precise as possible. For instance, the momentum distribution could be adjusted to better fit the real data at low $\pT$. Or the entire thermal model could be replaced by mixed events created from real data~\cite{Adamczyk:2017yhe}.

For the training, the supervised learning techniques that are applied need a truth value assigned to each sample, i.e.\ to each jet.
In the present model, the truth that will be approximated by the new correction method is the true jet momentum.
Here, it is defined as the reconstructed jet momentum multiplied by the momentum fraction that is carried by PYTHIA particles in the jet,

\begin{equation}
  \jettrue = \jetraw \cdot \sum_i p_{\mathrm{T,\;const}\;i}^\mathrm{PYTHIA}/ \sum_i p_{\mathrm{T,\;const}\;i}.
\end{equation}

In this equation, \jetraw\ is the reconstructed jet transverse momentum before any background correction. 
With this definition of the true jet momentum, also the true background is implicitly defined: It consists of all the particles from the thermal model.
As an alternative definition, the true jet momentum can also be defined as the momentum of jets reconstructed by PYTHIA particles only and geometrically matched to the reconstructed jets in the full toy event.
Since the background influences the jet finding algorithm, these matched jets are conceptually closer to the perfectly corrected jets. On the other hand, this definition is technically more complex since it needs further parameters like the matching radius and is thus less robust. There is a chance for mismatching jets, in particular at low transverse momentum.
However, in the end both truth definitions allow to train a model with very similar performance.

The toy model dataset is split into a training and a testing dataset, which is a standard procedure for ML-based approaches.
In total, $5$M toy events were generated from which roughly $6.5$M samples were extracted for further analysis. More samples would be available in the generated events, but only fractions are used to save memory. In subsequent analyses, these fractions are fully taken into account in the normalization.
For the training, $10\%$ of the data is used. The remaining $90\%$ form the testing dataset.

For the training of the estimators, only jets with $\jettrue > 5$ \GeVc\ are processed. The inclusion of jets below this threshold in the training distorted the corrected spectrum at higher \pT. The jet production in bins of $p_\mathrm{T,\,hard}$ have not been reweighted before being used for the training. Therefore, the training sample consists of more jets at higher transverse momentum compared to the natural abundances. This has the advantage that the training sample is not dominated by low-\pT\ jets.

Besides the evaluation of the background estimator concerning its resolution and potential biases, the performance of the correction method for actual heavy-ion-like collision data is also tested. For this purpose, the toy model cannot be used as it represents just one hard jet--jet interaction in a soft background. Instead, a test dataset of $2$M events is created by HIJING at the energy of $\snn =$ 2.76 TeV, with jet quenching switched off in the simulation. HIJING is tuned to roughly reproduce the expected multiplicities. An impact parameter of $b < 3.6$ fm and $p_\mathrm{T,\;hard} > 5.8$ leads to a nearly Gaussian multiplicity distribution with a mean value of $N \approx 2200$ and a width of 200, which is further discussed below.

\section{Background estimators}
As introduced above, the established method to correct the transverse momentum of jets is based on a per-event background estimate~\cite{Cacciari:2007fd}.
In each event, clusters are reconstructed using the $k_\mathrm{T}$ algorithm because it is particularly sensitive to background.
After removing the two hardest clusters from the cluster collection, the estimated background density is calculated as the median of the $k_\mathrm{T}$-cluster momentum densities.
It is defined by
\begin{equation}
  \rho = \mathrm{median} \left(\frac{p_{\mathrm{T},\;i}}{A_i}\right),
\end{equation}
where $i$ represents the index of all $k_\mathrm{T}$-clusters of the collection and $A_i$ their areas.
The median provides stability against outliers in the underlying distribution, e.g.\ against outliers from the jet signal clustered by the $k_\mathrm{T}$ algorithm.
Jets are corrected for this background density on a jet-by-jet basis by subtracting the background density multiplied by the jet area, $\jetrec= \jetraw - \rho \cdot A$.

Residual fluctuations are usually treated in an unfolding procedure.
They can be quantified by embedding jets of known transverse momenta into heavy-ion events. 
These jets are affected by the heavy-ion environment and, therefore, their reconstructed momenta are directly modified by residual fluctuations.
$\delta\pT$ quantifies the fluctuations and is defined by
\begin{equation}
\label{dptembedding}
  \delta\pT = \jetrec - \jettrue,
\end{equation}
for reconstructed jets matched with a known reference jet.
As described in detail in \cite{Abelev:2012ej}, these fluctuations can be approximated by a $\Gamma$-function.
In an alternative procedure to quantify residual fluctuations, one calculates the background-corrected momentum of particles in jet-sized cones randomly placed in heavy-ion events.\\

\begin{figure}[tbh!]
\begin{center}
  \includegraphics[width=1.0\textwidth]{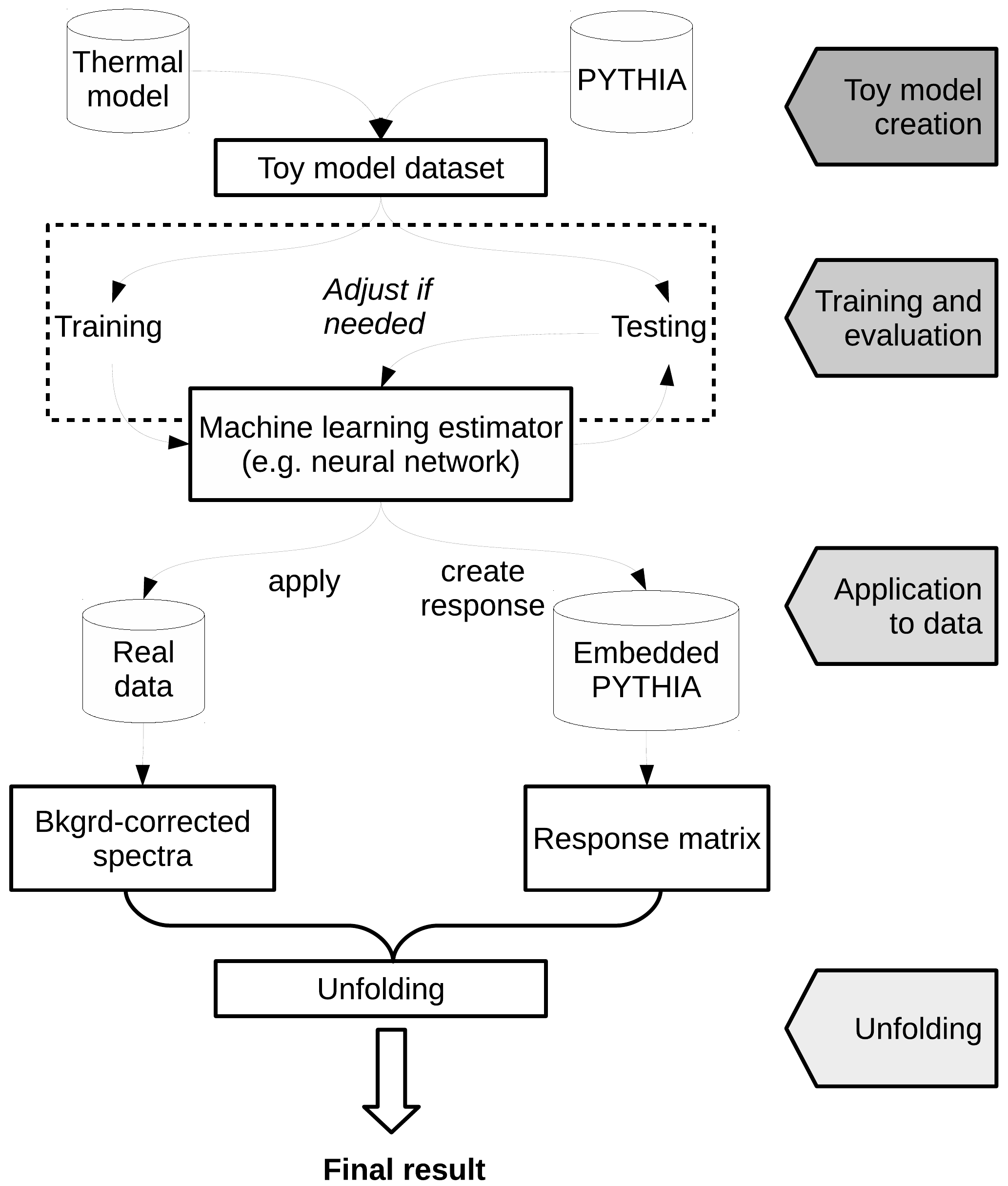}
  \caption{Illustration of the full proposed analysis strategy. Explanations are given in the text.}
  \label{fig:analysisFlow}
\end{center}
\end{figure}

The novel background estimator introduced in this paper follows a different approach. 
Instead of calculating the background once for an event, the background correction is applied completely on a jet-by-jet basis.
The main idea behind this is that the properties of background particles and those belonging to the jet process are very different.
While the background is dominated by low-$\pT$ particles, the jet signal should contain sizable higher momentum contributions.  
The information on the fraction of momentum from background is therefore partially encoded within the jet.
However, the relation of the input parameters, including jet constituents, and the true jet momentum is not trivial.
Machine Learning techniques are ideal to approximate this mapping by learning from data instead of modeling the relation by hand with expert knowledge. 
Since the task is to predict a numerical value for each sample, a regression algorithm needs to be used.

Several algorithms have been evaluated and will be compared in the next section: shallow neural networks~\cite{Haykin1998}, random forests~\cite{Breiman2001}, and linear regression.
The neural network is implemented as a shallow multi-layer perceptron, consisting of three layers with [100, 100, 50] nodes. The ADAM optimizer~\cite{KingmaB14}, an improved stochastic gradient-based optimizer, is used and the nodes/neurons are activated by the ReLU activation function~\cite{NairHinton2010}.
For the random forest algorithm, an ensemble of 30 decision trees is used. The linear regression model is the simplest model probed.
All models are used as implementations within the {\it scikit-learn}~\cite{scikit-learn} Python module. If a parameter is not explicitly given above, its default setting in {\it scikit-learn} v0.19.2 is used.

In order to find a suitable combination of input parameters, the analysis was repeated for a large variety of parameter sets. The number of parameters used is kept small to avoid a dependence on data subtleties. Eventually, the following input parameters prove to be useful, discriminative features:\\
(1) The uncorrected jet momentum as reconstructed by the jet finding algorithm,\\
(2) the jet transverse momentum, corrected by the established area-based method,\\ 
(3) several jet shape observables, namely jet mass, radial moment, momentum dispersion, and LeSub,\\
(4) the number of constituents within the jet,\\
(5) mean and median of all constituent transverse momenta,\\
(6) the transverse momenta of the first ten leading, i.e.\ hardest, particles within the jet.

The new ML-based approach defines a background estimator that works on a jet-by-jet basis and that uses the jet constituents as major input. Residual fluctuations -- which turn out to be smaller than for the established area-based background correction method -- can still be corrected in an unfolding procedure. The corresponding response should be estimated with the embedding method, given in Eq.~\ref{dptembedding}.\\


The proposed analysis strategy to train, validate, and apply the model to real data is depicted in Fig.\ref{fig:analysisFlow}.
The first step is the creation of the toy model data, which is described in the previous section.
Proton--proton data promise to provide more realistic vacuum jets compared to PYTHIA simulations and can be used instead if available for comparable detector conditions and same energy.
For the same reason, as a more accurate alternative to a thermal model, the vacuum jets can also be embedded into heavy-ion events or mixed heavy-ion events.
Background estimators trained on different toy model configurations could serve for a systematic uncertainty analysis on real data.
The second step is the training of the ML-based estimators and their evaluation on the toy model as will be described in the next section.
Training and evaluation dataset are independent subsamples of the full toy model dataset.
During this step, the model hyperparameters can be adjusted to obtain a good performance.
These first two steps are presented in this paper.
In the third step, the trained and commissioned estimator is applied to heavy-ion data to obtain background-corrected spectra, e.g.\ jet momentum or jet mass spectra.
In addition, a response matrix is created to unfold residual fluctuations (and possible detector effects).
We propose to create the response matrix by embedding vacuum jets into a realistic background, e.g.\ given by data.
The last step is the unfolding procedure to measure the final spectra.\\

\section{Comparison of the background estimators}
To assess the performance of the new ML-based background estimator, it is applied to data created by the PYTHIA + thermal toy model introduced above and compared to the established area-based method.
A direct measure for the resolution and the overall bias of an estimator is the distribution of residuals $\jetrec - \jettrue$. The residual distributions are normalized to unity to present a probability density and shown with Gaussian fits.
The mean of these fits can be interpreted as the overall bias, where the width $\sigma$ indicates the resolution of the method.

In the following, some results will only be presented for the neural network estimator. However, in these cases, the other estimators exhibit qualitatively the same results.

\begin{figure}[tbh!]
\begin{center}
  \includegraphics[width=1.0\textwidth]{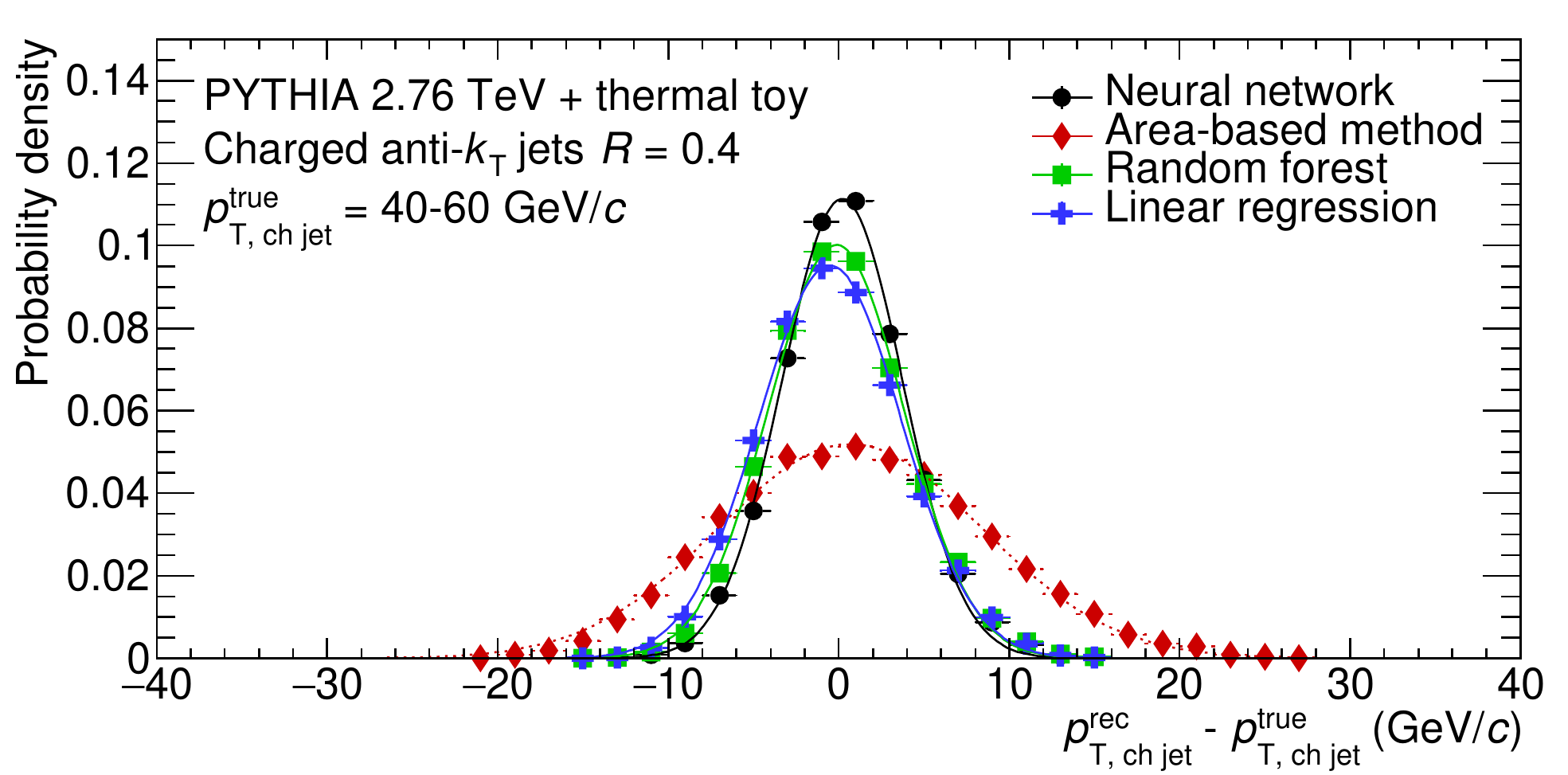}
  \caption{Residual distributions for several background estimators and $40 \leq \jettrue < 60$ \GeVc.}
  \label{fig:ResidualDistributions1}
\end{center}
\end{figure}

In Fig.~\ref{fig:ResidualDistributions1}, the residual distributions of the estimators considered are shown for jets with $40 \leq \jettrue < 60$ GeV/$c$.
It is immediately seen that all ML-based estimators considered exhibit a superior performance compared to the area-based correction method. While, to first order, the ML-based estimators all perform equally well on the toy model, the area-based correction has an inferior resolution, as summarized in \Tab{tab:residuals}.
The width of the residuals is a factor two smaller for the new background method.
These smaller fluctuations also lead to much less purely combinatorial jets at low jet transverse momentum and, therefore, allow to measure jets down approximately factor two lower \pT.

\begin{figure}[tbh!]
\begin{center}
  \includegraphics[width=1.0\textwidth]{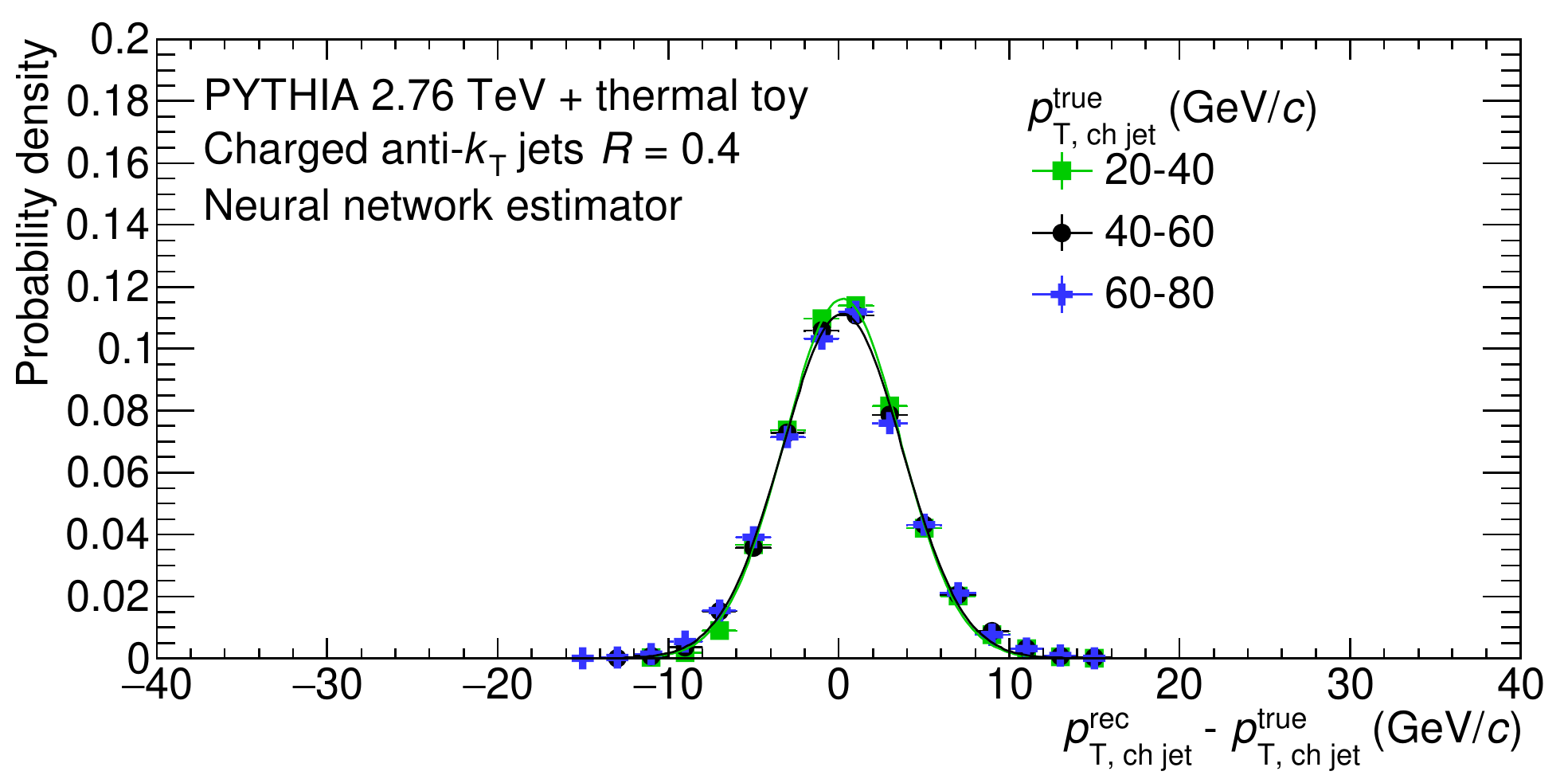}
  \caption{Residual distributions for the neural network estimator and several (true) jet transverse momenta.}
  \label{fig:ResidualDistributions2}
\end{center}
\end{figure}

It can also be shown that the new background estimator works similarly for different $\jettrue$.
The $\jettrue$-dependence of the estimator is depicted in Fig.~\ref{fig:ResidualDistributions2}.
The estimator has roughly the same resolution and no sizable bias.

\begin{figure}[tbh!]
\begin{center}
  \includegraphics[width=1.0\textwidth]{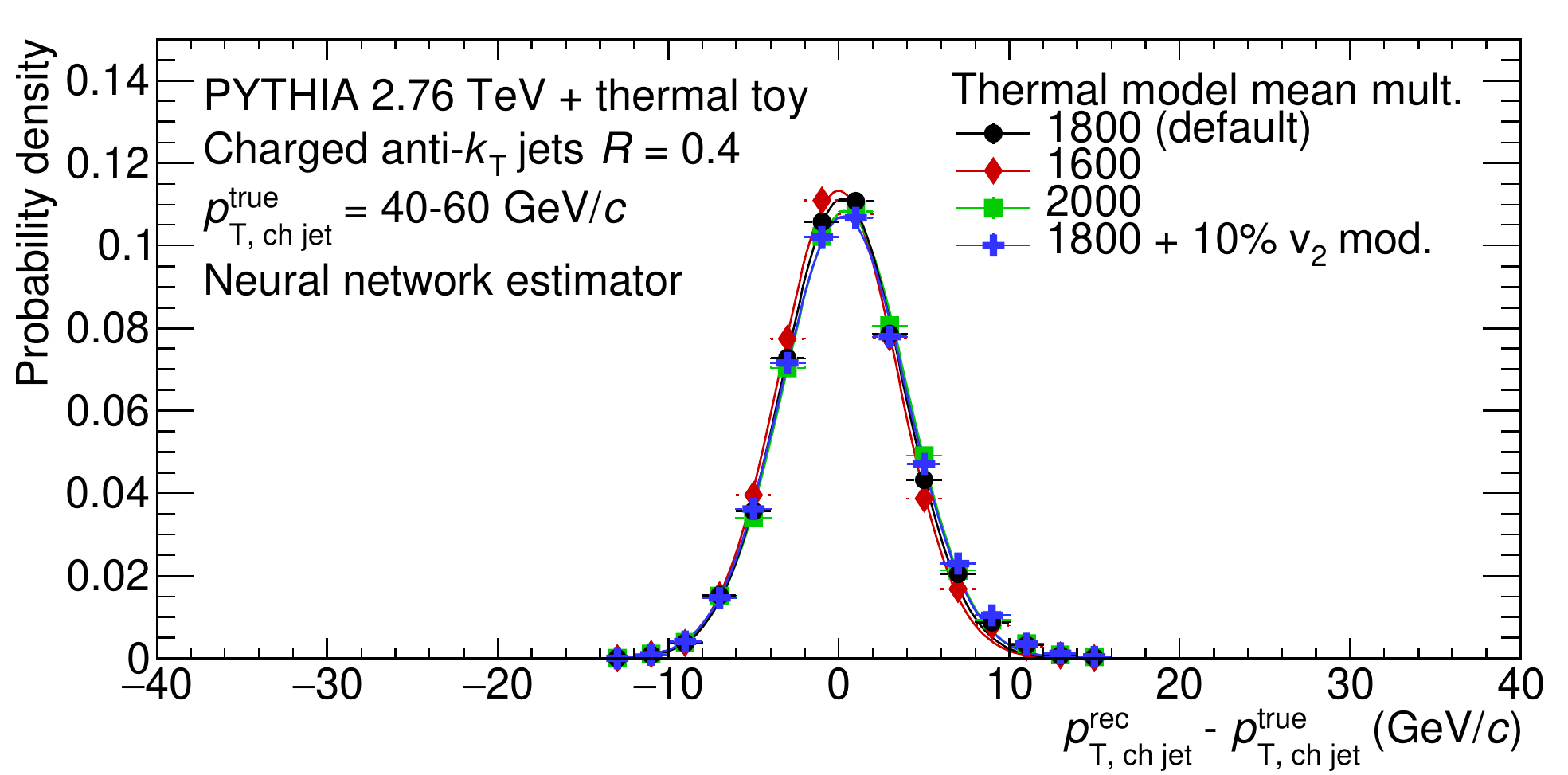}
  \caption{Comparison of residual distributions for several thermal background settings for $40 \leq \jettrue < 60$ \GeVc.}
  \label{fig:ResidualDistributions_Mult_NN}
\end{center}
\end{figure}
\begin{figure}[tbh!]
\begin{center}
  \includegraphics[width=1.0\textwidth]{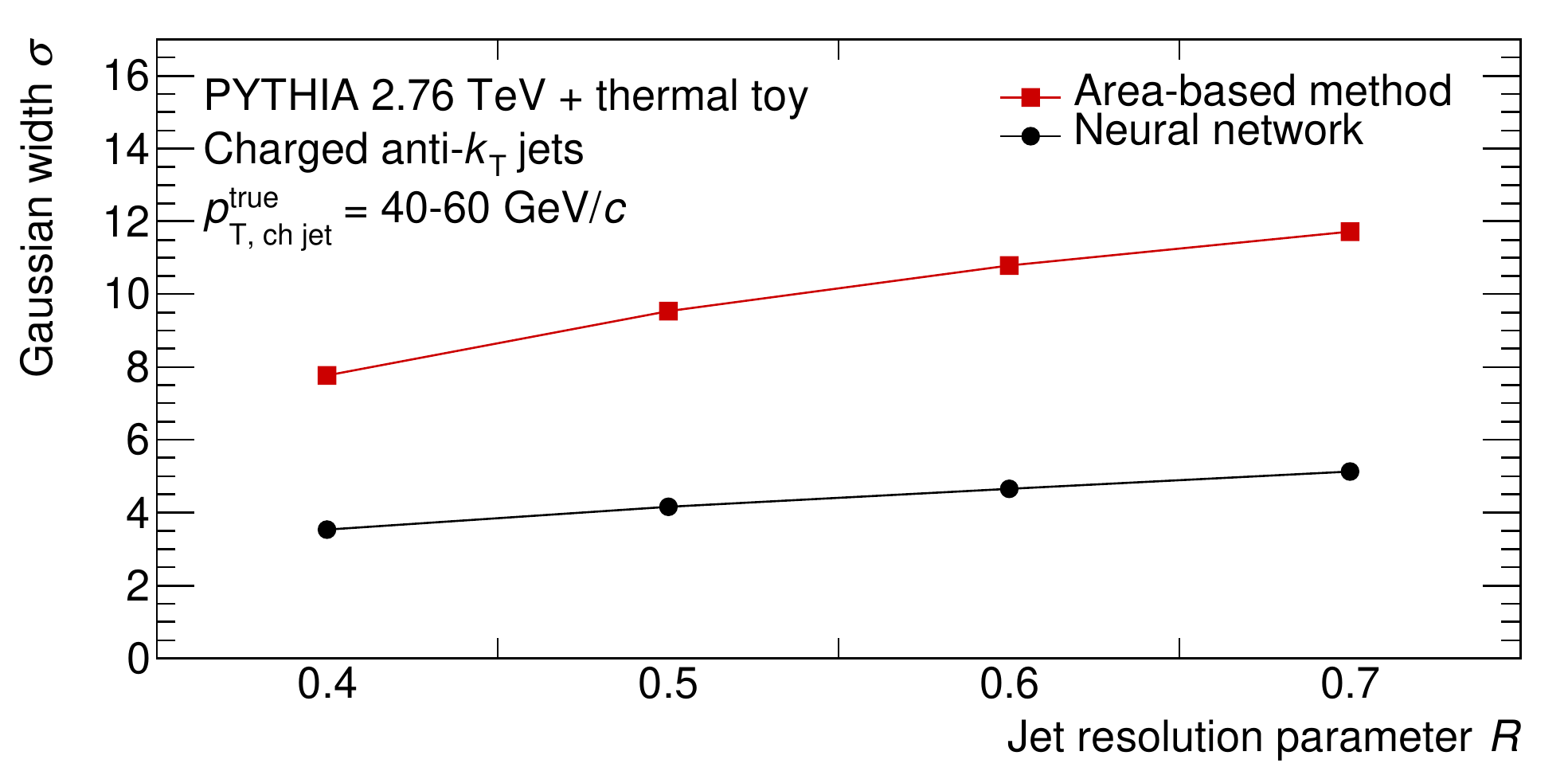}
  \caption{Comparison of the widths of the residual distributions as a function of the jet resolution parameter for $40 \leq \jettrue < 60$ \GeVc.}
  \label{fig:RadiusDependence_NN}
\end{center}
\end{figure}

An important property of the estimator is how robust it is to modifications in the underlying background or with respect to jet fragmentation or jet type. The estimators performance should not  strongly depend on these ingredients of the toy model. In reality, the background is definitely more complex than in the present model and the jets will be different in real data.
First, the jets in the toy model are generated by PYTHIA and the PYTHIA-simulated jet fragmentation will be slightly different than in real data.
Second, the fraction of quark and gluon jets is different in heavy-ion collisions than in PYTHIA, so a dependence on the jet type could bias the results.
And finally, for the same reason, the background estimator should not yield qualitatively different results for quenched and unquenched jets, which may differ in shape and fragmentation.

To estimate the sensitivity to differences in the underlying thermal background, the correction is applied in events with modified thermal background multiplicities. The multiplicities are adjusted by roughly 10\%, from the baseline of $N=1800$ to $N=1600$ and $N=2000$.
Hydrodynamic particle flow is not part of the thermal toy model but will occur on data.
Its influence on the model performance is checked by applying the model to a particle distribution with $N=1800$ which is modulated with $v_2 = 0.1$ in azimuth $\phi$ using
\begin{equation}
  \mathrm{d}N_\mathrm{flow}/\mathrm{d}\phi = \mathrm{d}N/\mathrm{d}\phi \cdot (1+ 2\cdot v_{2} \cos(2\phi)).
\end{equation}
This corresponds to a relative particle modulation of up to 20\% in the azimuthal yield distribution.
It turns out that these changes have only a negligible effect on the model performance, as seen in Fig.~\ref{fig:ResidualDistributions_Mult_NN}. Here, the impact of different thermal backgrounds is presented for $40 \leq \jettrue < 60$ \GeVc, but also for lower jet transverse momenta it is very small.

Since jets are extended objects consisting of correlated particles up to large distances far beyond $R=0.4$, using a large jet resolution parameter is in general desirable.
While the presented study was performed for jets with $R=0.4$, the performance of larger jet resolution parameter was also tested.
\Figure{fig:RadiusDependence_NN} presents the dependence of the background estimator on the jet resolution parameter as the $R$-dependence of the width of the residual distribution, obtained by a Gaussian fit. As expected, the width gets larger, i.e.\ the resolution worsens, for larger $R$. However for the ML-based approach, both, the increase with increasing $R$, as well as the absolute value of the width, are significantly smaller than for the area-based correction.

\begin{table}
\centering
\begin{tabular}[t]{l|ll|ll|ll}
Estimator & \multicolumn{2}{c}{Inclusive jets} & \multicolumn{2}{c}{Quark jets} & \multicolumn{2}{c}{Gluon jets}\\ \hline
Fit param.\ (GeV/$c$)  & Mean & $\sigma$ &  Mean & $\sigma$ &  Mean & $\sigma$ \\ \hline
Area-based method    & 0.7 & 7.8 & 0.8 & 7.3 & 0.9 & 7.7\\
Neural network    & 0.2 & 3.5 & 1.0 & 3.1 & -0.5 & 3.6\\
Random Forest     & -0.1 & 4.0 & 0.7 & 3.6 & -0.8 & 3.9\\
Linear regression    & -0.5 & 4.2 & 0.8 & 3.8 & -1.4 & 4.1\\
\end{tabular}
\caption{\label{tab:residuals} Properties of the residual distributions for the considered estimators for $40 \leq \jettrue < 60$ \GeVc. Mean and $\sigma$ represent mean and standard deviation of the Gaussian fits, respectively.}
\end{table}

\begin{figure}[tbh!]
\begin{center}
  \includegraphics[width=1.0\textwidth]{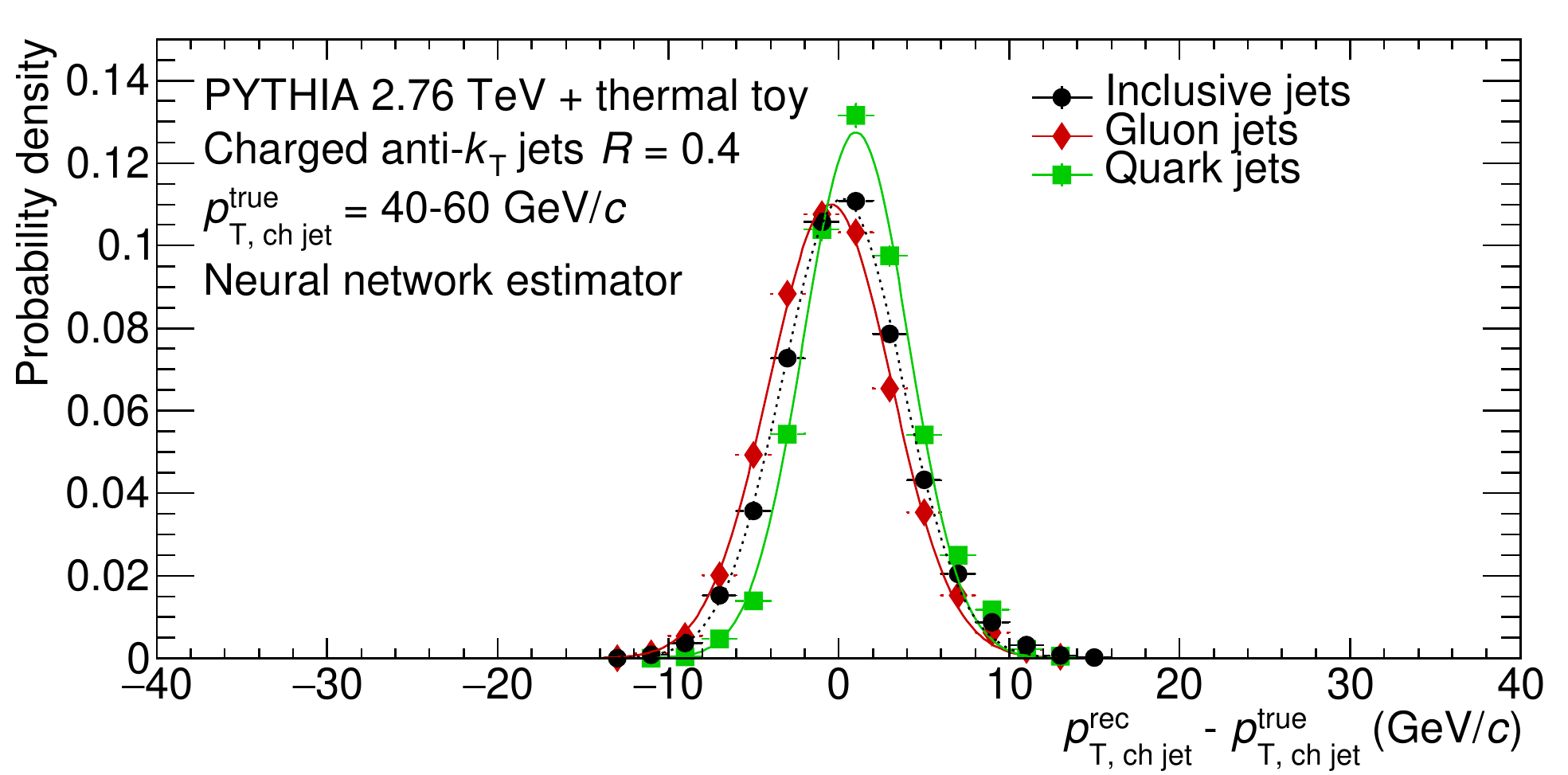}
  \caption{Comparison of residual distributions for gluon, quark, and inclusive jets for the neural network estimator.}
  \label{fig:ResidualDistributionsQG1}
\end{center}
\end{figure}

\begin{figure}[tbh!]
\begin{center}
  \includegraphics[width=1.0\textwidth]{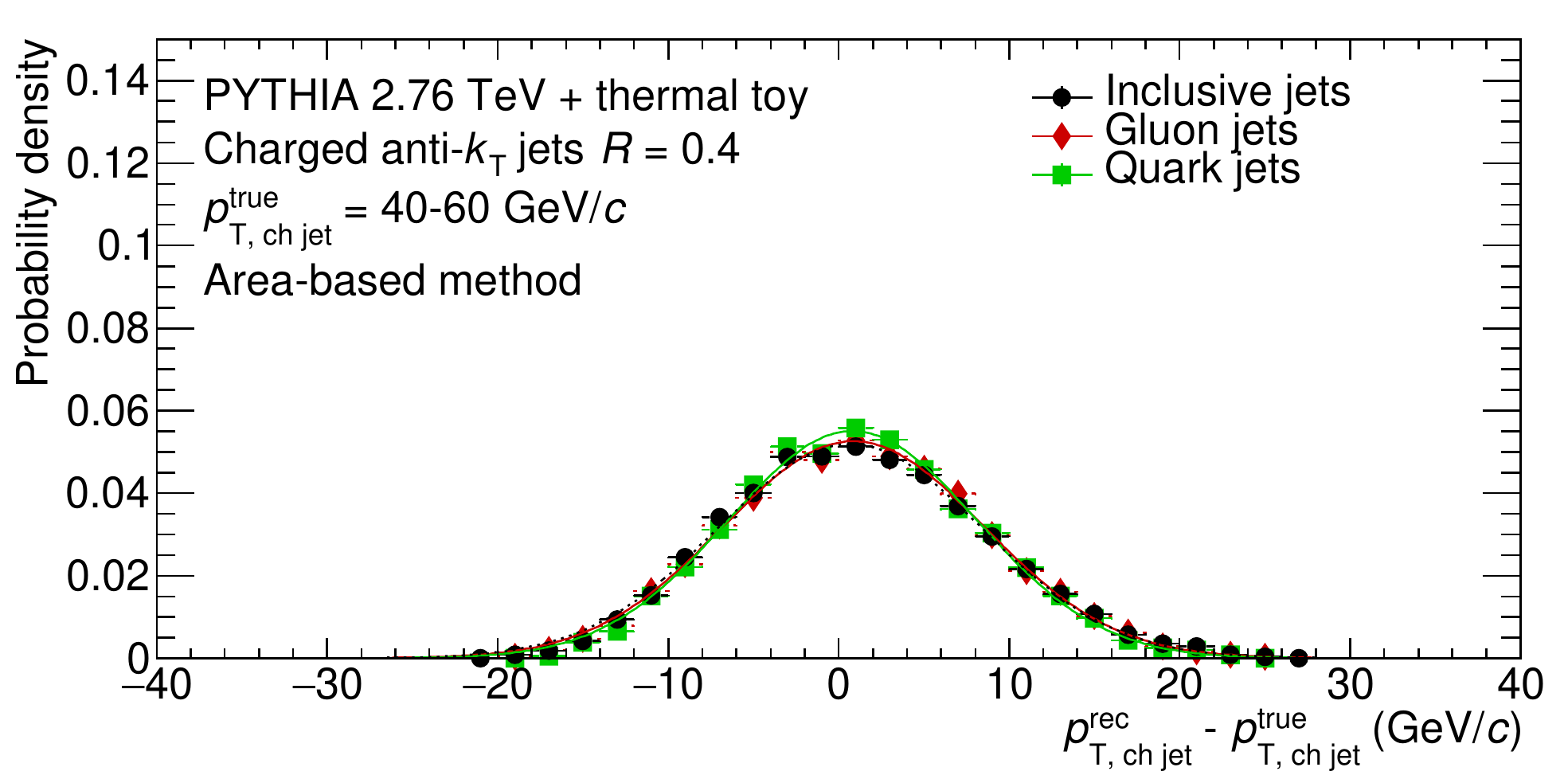}
  \caption{Comparison of residual distributions for gluon, quark, and inclusive jets for the area-based method.}
  \label{fig:ResidualDistributionsQG2}
\end{center}
\end{figure}

\begin{table}
\centering
\begin{tabular}[b]{ll|ll}
Feature & Score  & Feature & Score \\ \hline
Jet \pT\ (no corr.) & \textbf{0.1355} & $p_\mathrm{T,\;const}^{1}$ &  0.0012\\
Jet mass & 0.0007 & $p_\mathrm{T,\;const}^{2}$ &  \textbf{0.0039}\\
Jet area & 0.0005 & $p_\mathrm{T,\;const}^{3}$ &  0.0015\\
Jet \pT\ (area-based corr.) & \textbf{0.7876} & $p_\mathrm{T,\;const}^{4}$ &  0.0011\\
LeSub & 0.0004 & $p_\mathrm{T,\;const}^{5}$ &  0.0009\\
Radial moment & 0.0005 & $p_\mathrm{T,\;const}^{6}$ &  0.0009\\
Momentum dispersion & 0.0007 & $p_\mathrm{T,\;const}^{7}$ &  0.0008\\
Number of constituents & 0.0008 & $p_\mathrm{T,\;const}^{8}$ &  0.0007\\
Mean of const. \pT & \textbf{0.0585} & $p_\mathrm{T,\;const}^{9}$ &  0.0006\\
Median of const. \pT & 0.0023 & $p_\mathrm{T,\;const}^{10}$ & 0.0007\\
\end{tabular}
\caption{\label{tab:importances} Random forest feature importances. A higher score corresponds to a higher importance of the feature. $p_\mathrm{T,\;const}^{i}$ is the transverse momentum of the $i^\mathrm{th}$-hardest particles. The four most important features are marked in bold face.}
\end{table}
To analyze the impact of different jet fragmentation or jet types on the background estimator, the residual distributions are calculated for different jet types.
As a quite extreme test, quark and gluon jets serve here as proxies for jets with different fragmentation. The residual distributions are presented in Figs.~\ref{fig:ResidualDistributionsQG1} and \ref{fig:ResidualDistributionsQG2}.
A small bias can be observed, which is of the order of the inclusive jet bias, see also Tab.~\ref{tab:residuals}.

\begin{figure}[tbh!]
\begin{center}
  \includegraphics[width=1.0\textwidth]{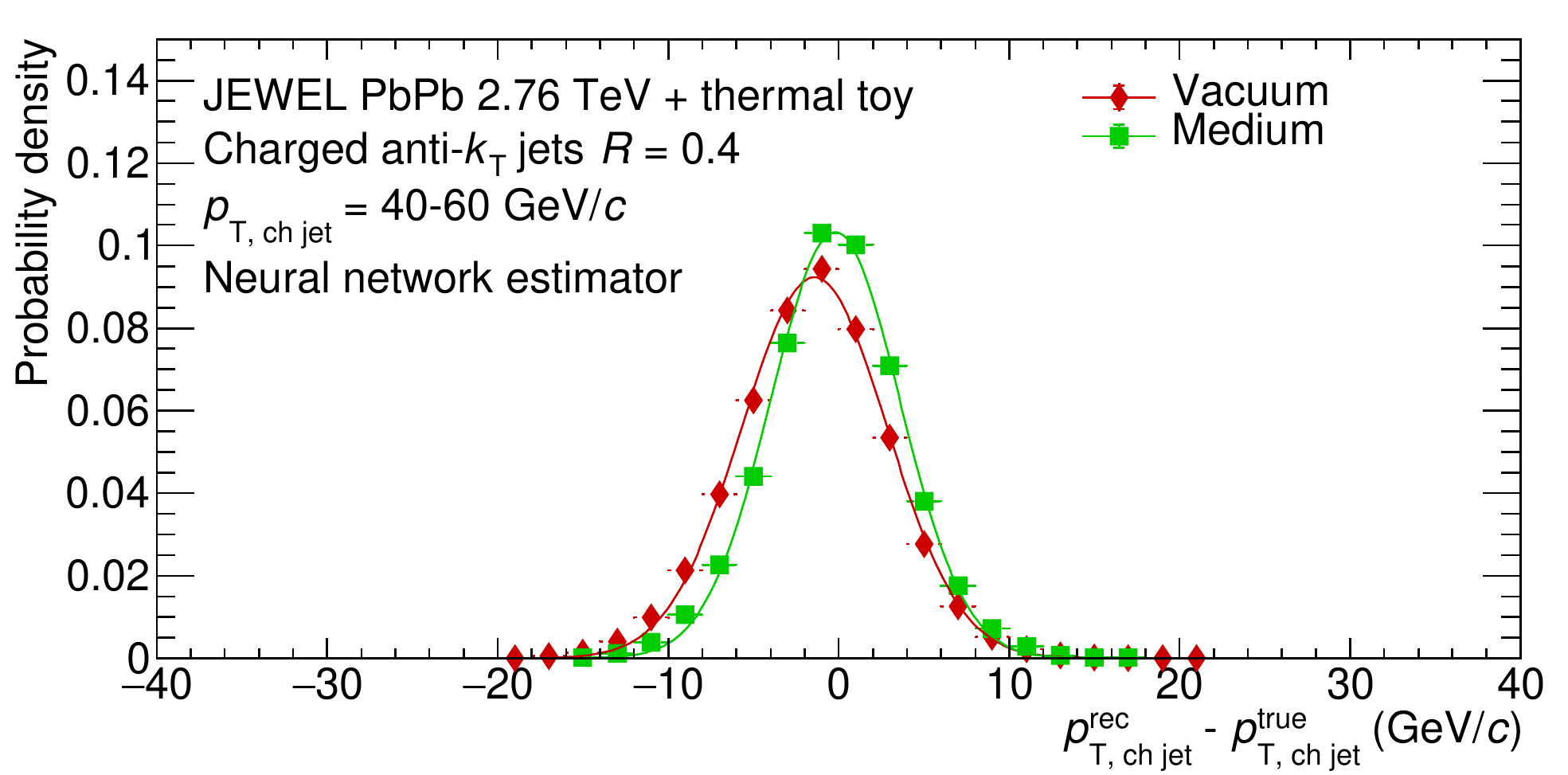}
  \caption{Comparison of residual distributions for jets created by JEWEL for the neural network estimator.}
  \label{fig:ResidualDistributionsJEWEL_NN}
\end{center}
\end{figure}

Another check on the fragmentation dependence of the background estimator was performed with jets generated by JEWEL~\cite{Zapp:2013vla} with the \textit{no recoils}-setting. JEWEL is a Monte Carlo generator that simulates the jet evolution within a medium created in heavy-ion collisions and was used to produce samples of jets affected or not affected by medium effects.
The residual distribution for the neural network estimator can be found in Fig.~\ref{fig:ResidualDistributionsJEWEL_NN}. Like for the quark/gluon jet comparison, a small bias can also be observed for JEWEL jets in the medium. The bias is of similar magnitude as the quark/gluon bias, though slightly smaller.

It should be emphasized that even a small bias can have a sizable effect on the jet production yields due to the steeply falling jet spectrum. It might lead to deviations of up to approximately 20\% at low transverse momentum around 20 \GeVc\ for the worst-case of pure quark or gluon samples. For more realistic jet collections, the bias will be significantly lower. In an analysis on real data, a systematic uncertainty should be evaluated taking this effect into account. Note though, that this bias will at least partially be corrected in an unfolding procedure.

\begin{figure}[tbh!]
\begin{center}
  \includegraphics[width=1.0\textwidth]{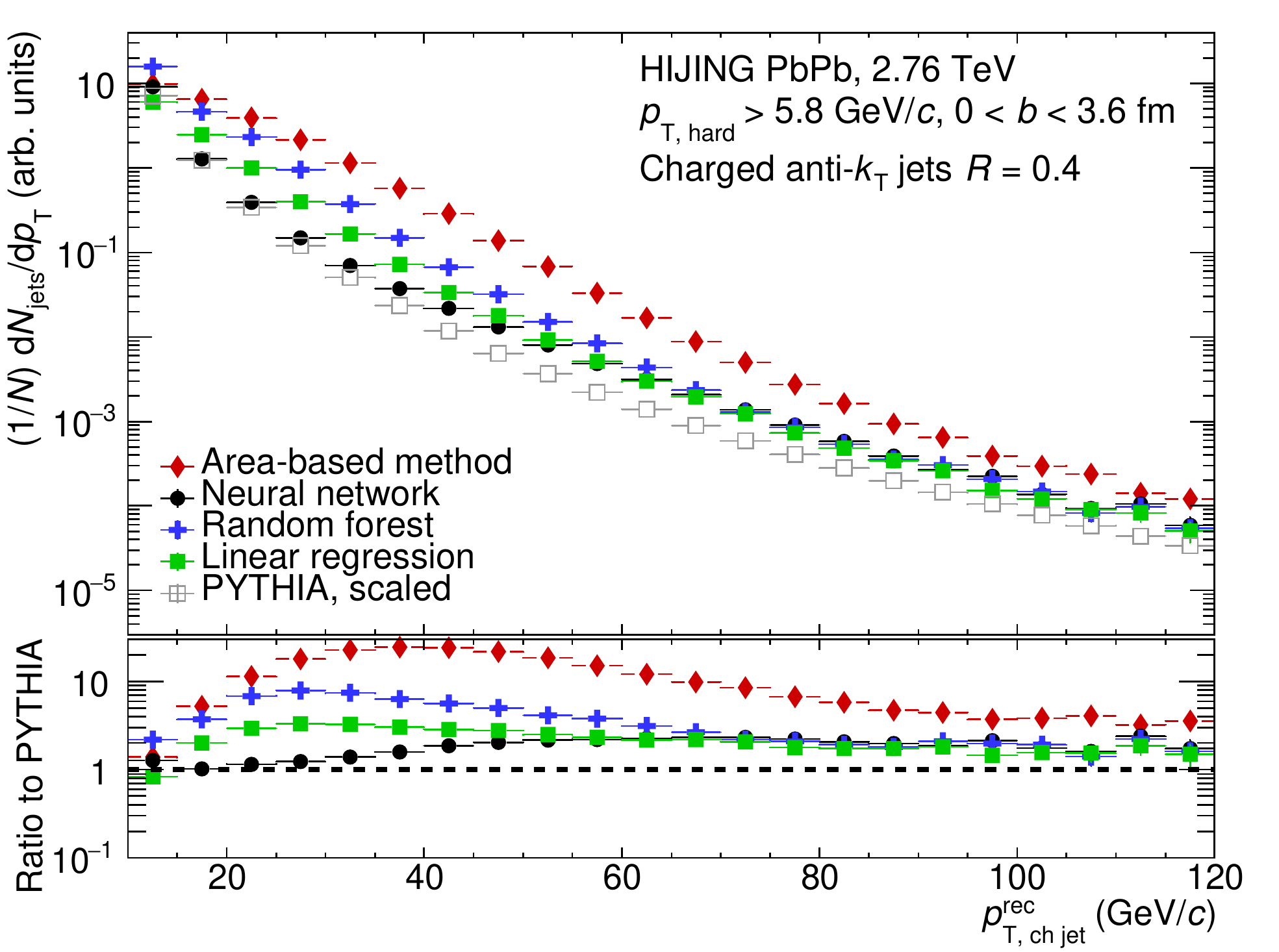}
  \caption{Reconstructed charged jet spectra in HIJING events and the ratio to ($N_\mathrm{coll}$-scaled) PYTHIA jet spectra.}
  \label{fig:Spectra_MethodComparison_HIJING}
\end{center}
\end{figure}

The random forest estimator in {\it scikit-learn} directly calculates the mean decrease in impurity in the tree ensemble, also called Gini importance~\cite{GiniImportance}, for each input parameter/feature.
In the Gini definition, a feature is more important when it is used in more tree splittings of more samples. 
Feature importances are simple measures (potentially) indicating how the random forest estimator learns from the data. Since all estimators perform approximately equally well, the importances should also give a rough estimate for the features' importances in the other estimators. The values can be found in Tab.~\ref{tab:importances}.
The most important features are all related to variables that characterize the ``hardness'' of the jet. The most important feature turns out to be the area-based corrected jet momentum, indicating that the area-based correction already gives a reasonable estimate for the true jet momentum. Interestingly, it is not the hardest particle within the jet which has the highest importance but the second-hardest.

The performance evaluation on the toy model data is helpful in gaining insights on how precisely the true momentum of a jet in a heavy-ion-like background is reconstructed by the correction method.
Unfortunately, the corrected jet spectra in the toy model are not realistic heavy-ion-like jet spectra, since they are made from one PYTHIA spectrum, embedded in thermal background. This changes drastically the amount of expected purely combinatorial jets as well as the jet spectra itself. Therefore, the toy model jet spectra are not presented here.

Instead, an independent study is performed on HIJING simulated data, shown in Fig.~\ref{fig:Spectra_MethodComparison_HIJING}. Central HIJING events are simulated and the new correction method~(without retraining) is applied as if the HIJING data were real data. The corrected jet spectra are then compared to $N_\mathrm{coll}$-scaled PYTHIA spectra to judge the performance of the background estimators.
The spectra comparison reveals that the new background estimators are also superior for realistic heavy-ion-like events down to very low transverse momenta of about 20 \GeVc. The spectra are closer to the expected truth and, therefore, an unfolding procedure of residual fluctuations will be possible with much higher precision and smaller uncertainties than for the standard method.
As a further cross check, several HIJING spectra were analyzed for different minimum $p_\mathrm{T,\;hard}$-cuts and different impact parameter ranges, leading to the same conclusion.

\begin{figure}[tbh!]
\begin{center}
  \includegraphics[width=1.0\textwidth]{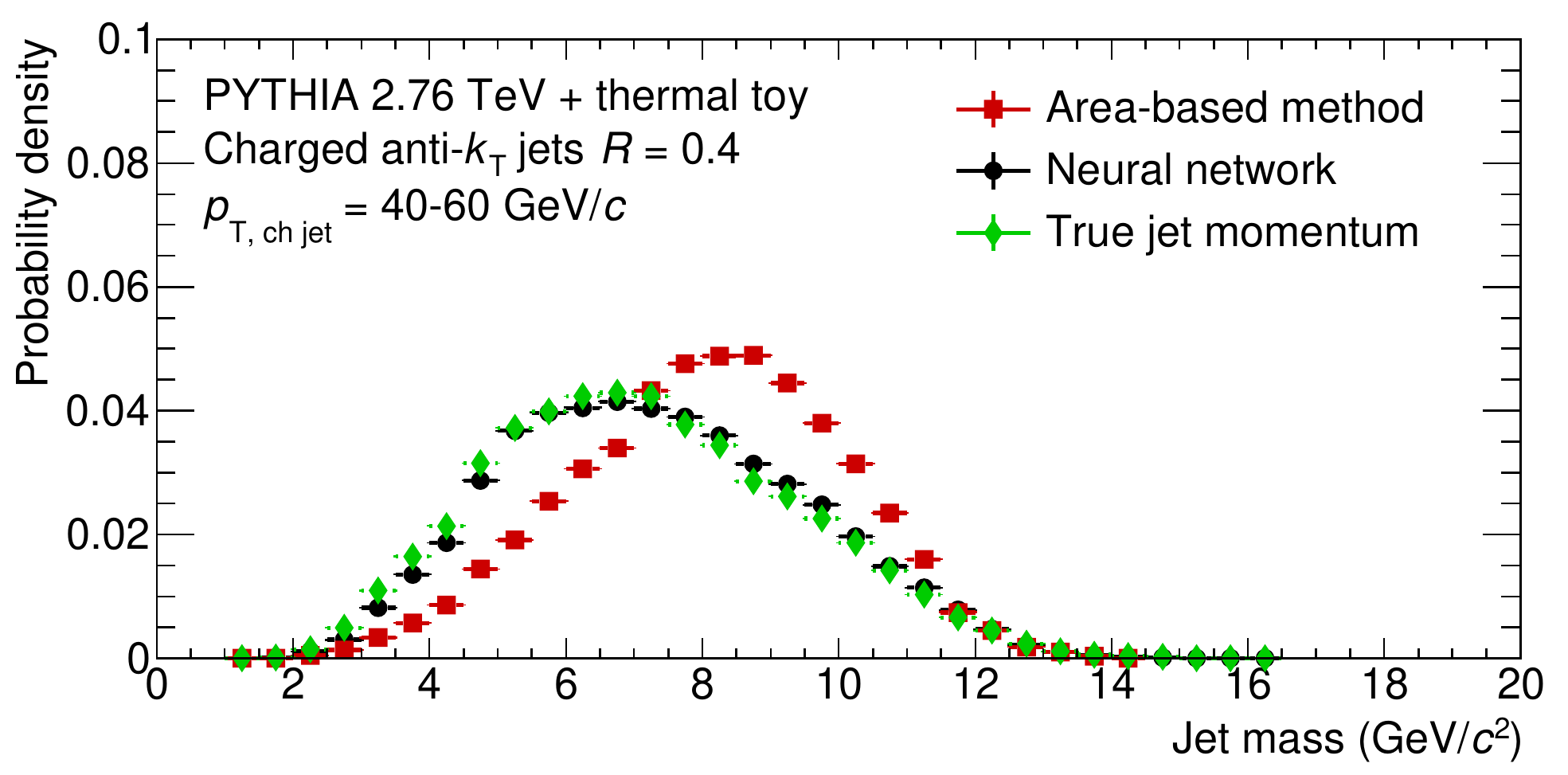}
  \caption{Jet mass distribution compared for different jet energy scales. A background correction for the jet mass itself was also applied~\cite{Soyez:2012hv}.}
  \label{fig:JetMassDifferentEnergyScales}
\end{center}
\end{figure}

In order to test how the measurement of the jet mass is affected by the different jet energy scales introduced by the background estimators, the jet mass has been measured for the toy model data for area-based and neural network background estimators.
Figure~\ref{fig:JetMassDifferentEnergyScales} shows the jet mass for jets with $40 \leq {p}_{\mathrm{T,\;ch\,jet}} < 60$ GeV/$c$, where the definition of $\jetrec$ -- and therefore the jet energy scale -- differs for the shown distributions.
The jet mass itself is also background-corrected~\cite{Soyez:2012hv}.
It turns out that the reconstructed jet masses coincide well for the true and ML-based jet energy scale. Therefore, the higher precision of the jet energy scale might allow a jet mass measurement at lower jet momenta than currently possible.
The jet mass for the area-based method energy scale seems to be significantly overestimated. Further tests show that these general observations are also valid for higher transverse jet momenta, although the bias of the area-based method decreases.

\section{Summary and outlook}
A novel method to correct for the large and strongly-fluctuating background in central heavy-ion collisions
using Machine Learning techniques is introduced.
In contrast to the established correction method, the background is estimated on a jet-by-jet basis rather than event-by-event.
A toy-model analysis indicates that the new background estimator is superior to the established method in terms of precision.
Studies also show that the residual fluctuations do not depend on the jet transverse momentum above $\jettrue \geq 20$ \GeVc.
The method was evaluated on several different underlying background models to investigate its susceptibility to differences compared to real data, 
revealing that the performance does not strongly depend on subtleties of the background.
In addition, quark and gluon jets as well as JEWEL-generated jets are used as proxies for a different jet fragmentation, with a small bias observed.
An analysis on real data should take into account a possible systematic uncertainty that could arise from this bias.
It has also been shown that the jet mass reconstructed with the ML-based jet energy scale coincides well with the mass in the true jet energy scale.
Hence, the application of the new method on real experimental data may allow considerably more precise measurements in heavy-ion collisions, in particular at low transverse momentum.

\section*{Acknowledgments}
\enlargethispage{1cm}
We would like to thank the ALICE Collaboration for providing computational resources for this study.
We are grateful to L.~Cunqueiro and J.~Harris for useful discussions and comments to the manuscript.
R.~Haake is supported in part by a Feodor Lynen Research Fellowship from the Alexander von Humboldt Foundation and by the U.S. Department of Energy, Office of Science, Office of Nuclear Physics under Grant number DE-SC004168. 
C.~Loizides is supported by the U.S. Department of Energy, Office of Science, Office of Nuclear Physics, under contract number DE-AC05-00OR22725.
\bibliographystyle{utphys}
\bibliography{biblio}
\end{document}